\documentclass[twocolumn,showpacs,preprintnumbers,prd,superscriptaddress,nofootinbib]{revtex4}
\usepackage{doi}
\usepackage{hyperref}
\hypersetup{
  colorlinks=true,        
  linkcolor=blue,         
  citecolor=cyan,         
}

\usepackage{bm}
\usepackage{latexsym}
\usepackage{dcolumn}
\usepackage{amsmath,amsfonts,amssymb}
\usepackage{amsthm}
\usepackage{color}
\usepackage{xcolor}
\usepackage{graphicx,float}
\usepackage{comment}
\usepackage{hyperref}
\usepackage[normalem]{ulem}
\usepackage{enumitem} 

\usepackage{pdfpages}
\usepackage{epstopdf}
\usepackage[utf8]{inputenc}

\def\be {\begin{equation}}
\def\ee {\end{equation}}
\def\bea {\begin{eqnarray}}
\def\eea {\end{eqnarray}}
\def\bc {\begin{center}}
\def\ec {\end{center}}
\def\bfg {\begin{figure}}
\def\efg {\end{figure}}
\def\bi {\begin{itemize}}
\def\ei {\end{itemize}}

\def\beq{\begin{equation}}
\def\eeq{\end{equation}}
\def\br{\begin{eqnarray}}
\def\er{\end{eqnarray}}
\newcommand{\eel}[1] {\label{#1}\end{equation}}

\newcommand{\bdm}{\begin{displaymath}}
\newcommand{\edm}{\end{displaymath}}

\usepackage{orcidlink}

\begin{document}

\title{Regular black holes with gravitational self-energy as dark matter }

\author{Kimet Jusufi\,\orcidlink{0000-0003-0527-4177}}
\email{kimet.jusufi@unite.edu.mk}
\affiliation{Physics Department, State University of Tetovo, Ilinden Street nn, 1200, Tetovo, North Macedonia}
\author{Douglas Singleton\,\orcidlink{0000-0001-9155-7282}}
\email{dougs@mail.fresnostate.edu}
\affiliation{Physics Department, California State University, Fresno, CA 93740}

\begin{abstract}
We incorporate the effect of non-local gravitational self-energy to obtain a neutral, non-singular spacetime geometry. This is achieved by using a non-local gravitational theory inspired by T-duality, where particle mass is not point-like but smeared over a region. 
This non-local gravitational self-interaction is derived from the Newtonian gravitational potential and energy density, allowing us to define a coordinate-independent quantity. 
Thus, we incorporate the non-local gravitational field into the spacetime metric. We demonstrate that the total ADM mass is modified by a finite, regularized gravitational mass term, leading to a regular solution of the Ayon-Beato-Garcia type metric but without electric charge. We show the existence of extremal configurations known as \emph{particle–black hole} objects of order of the Planck mass, which are thermodynamically stable, have a vanishing Hawking temperature and could be a viable dark matter candidate.
\end{abstract}

\maketitle
\section{Introduction}
Despite the remarkable success of general relativity in matching predictions with observations, it still faces a major limitation at small-distance or large-energy scales.  Specifically, various black hole and cosmological solutions of general relativity have singularities in their space-time metric. The theory breaks down at these singularities \cite{Penrose:1964wq,Hawking:1970zqf}.  
To address the problematic short-distance behavior of black hole space-times, some theories propose limiting space-time curvature as one approaches the Planck scale. One of the first models in this direction is the Bardeen black hole \cite{bardeen1968}, which was obtained by introducing nonlinear electrodynamics and magnetic charge to regularize the singularity \cite{Ayon-Beato:2000mjt}. There are other models that avoid the singularity via electric charge  \cite{Ayon-Beato:1998hmi}. These models rely either on undetected magnetic monopoles or electric charge, which tends to vanish rapidly due to Hawking radiation and Schwinger instability pair production \cite{Gibbons:1975kk}.

Meanwhile, several alternative models for non-singular neutral black holes have been developed~\cite{Dymnikova:1992ux,Hayward:2005gi}. A significant advance came from a class of regular black holes inspired by string theory, where noncommutative geometry is used to smooth out spacetime at small scales \cite{Nicolini:2005vd}. These models not only avoid singularities but also predict an end to black hole evaporation where one ends up with a zero temperature remnant with charge or spin.

T-duality plays a central role in motivating the emergence of a minimal length scale in quantum gravity. At its core, T-duality is a symmetry of string theory that relates compactifications on circles of radius $R$ and $R^\star = \alpha'/R$, effectively exchanging large and small distance scales. This duality implies that physics at distances smaller than the self-dual scale $R^\star$ is equivalent to physics at larger scales, thereby introducing a natural ultraviolet cutoff.

In the path-integral formulation, this T-duality manifests as an invariance under the transformation of the proper time parameter $s \rightarrow L^2/s$, leading to a modified propagator with an effective zero-point length. As a consequence, short-distance divergences are regulated, and spacetime acquires a non-local structure. This provides the theoretical basis for incorporating smeared sources and non-local gravitational effects in the present work.
In more detail, in the Schwinger representation of a field propagator, this T-duality implies invariance in the path integral of a particle of mass $m$, assigning the weight $\exp\left(-m(s +L^2/s)\right)$, which ensures equal treatment of paths with proper time $s$ and $L^2/s$ \cite{Nicolini:2022rlz,Nicolini:2023hub}, where $L$ is a minimal length. The resulting quantum-corrected propagator remains regular, incorporating an ultraviolet cut-off dictated by the zero-point length $l_0$ which is assumed to equal $2 L$ \cite{Padmanabhan:1996ap,Smailagic:2003hm,Fontanini:2005ik}. Further, the duality principle of the zero-point length of spacetime is related to the generalized uncertainty principle \cite{Jusufi:2023tdn} \cite{Mondal:2020zet} as well as the dimensional reduction of spacetime with zero-point length \cite{Mondal:2021izm}. 
 
The T-duality-inspired approach has been successfully applied to black hole physics, with exact solutions reported for both neutral and charged black holes \cite{Nicolini:2019irw, Gaete:2022ukm}.
Further, T-duality black holes have recently been studied in the context of the black hole singularity theorem. By applying the Raychaudhuri equation, it was shown that the presence of an initial trapped surface does not necessarily lead to a curvature singularity. Instead, quantum corrections can counteract gravitational collapse and regularize the spacetime~\cite{Jusufi:2024dtr}. Along these lines, T-duality-inspired cosmological models free from initial singularities have also been proposed~\cite{Jusufi:2022mir,Millano:2023ahb}.
 
 Here we study the effect of self-interaction gravitational energy in the spacetime of T-duality black holes. This effect was not included in the original derivation of T-duality black holes \cite{Nicolini:2019irw}. However, a similar effect was obtained for charged black holes in T-duality \cite{Gaete:2022ukm}. Here we show that using the self-interaction gravitational energy in the Newtonian limit one can indeed obtain a completely neutral and regular spacetime of the Ayon-Beato-Garcia type solution, without any charge and characterized by two parameters: mass and the zero point length. This solution is of particular interest since such a black hole is neutral and it does not suffer from the Schwinger instability compared to regular but charged metrics derived
by coupling nonlinear electrodynamics theory to gravity. From the thermodynamic point of view, such black holes can lead to stable remnant objects with vanishing Hawking temperature. As we show, this opens the possibility to address the problem of dark matter using these regular black holes \cite{Calza:2024fzo,Calza:2024xdh,Calza:2025mwn,Davies:2024ysj,Carr:2025auw,Asmanoglu:2025agc}. Most importantly, we report corrections to the total mass due to the non-local effect of the gravitational field. This relation has an extra mass term that can lead to an effect similar to dark matter. A key finding in the present work is the new regular black hole solution from non-local gravitational energy stored in the field. The solution is similar to the Ayon-Beato-Garcia spacetime, where the role of charge is played by the zero point length. 

 The paper is organized as follows. In Section II, we put forward a non-local T-duality gravitational theory and the Einstein field equations in this model.  In Section III we obtain the black regular hole solution with non-local gravitational effect. In Section IV, we elaborate the existence of black hole remnants and the possible contribution of such remnants to dark matter. In Section V, we comment on our findings. Throughout the work, we set $G=\hbar=c=1$, unless otherwise specified.

\section{Non-local gravitational self energy in T-duality}

In order to include the effect of gravitational energy stored in the field,  first we shall review the quantum-corrected static interaction potential according to the field theory having the path integral duality proposed in \cite{Nicolini:2019irw}. We recall that the momentum space massive propagator induced by the path integral duality is given by \cite{Nicolini:2022rlz,Nicolini:2023hub}
\begin{equation}
G(k)
= -\frac{l_0}{\sqrt{k^2+m_0^2}}\, K_1 \!\left(l_0 \sqrt{k^2+m_0^2}\right)
\end{equation}
with $l_0$ being the zero-point length and $K_1\!\left(x\right)$ is a modified Bessel function of the second kind. Without loss of generality, we consider the massless propagator case, i.e. $m_0=0$. 
Specifically, we have two cases: at small momenta, we obtain the conventional massless propagator $G(k) = -k^{-2}$. At large momenta, the exponential suppression is responsible for curing UV divergences \cite{Nicolini:2022rlz,Nicolini:2023hub}.

Consider a static external source $J$ which consists of two point-like masses, $m$ and $M$, at relative distance $\vec{r}$, then the potential is given by \cite{Nicolini:2019irw}
\begin{eqnarray}\notag
V_G(r)
&=& -M\, \int\!\frac{d^3 k}{{\left(2\pi\right)}^3}\; { G_{F}(k)|}_{k^0=0}\; 
 \exp\!\left(i \vec{k} \vec{r}\right) \\
 &=&
 -\frac{M}{\sqrt{r^2 + l_0^2}}. \label{potential}
\end{eqnarray}
In this section, we show how to include gravitational self-energy in a regular space-time under T-duality in the Newtonian limit. The problem of defining a local gravitational field energy is well known. This is due to the
equivalence principle and the non-local nature of gravitational energy in general relativity. Attempts to define gravitational energy locally lead to the introduction of energy-momentum pseudotensors, as proposed by Einstein \cite{Einstein:1916vd}, Landau-Lifshitz \cite{Landau:1975pou}, Papapetrou \cite{Papapetrou:1948jw}, Weinberg \cite{Weinberg:1972kfs} etc. However, all these proposals depend on the choice of coordinates and are not true tensors, making them unsatisfactory for a fully general definition. We will show that gravitational self-energy can arise from non-local geometric effects in the weak gravity regime, which we propose persists into the strong gravity regime. At the level of the field equations, we use the Einstein-Hilbert action
\begin{eqnarray}
    S_{\rm EH}=\frac{1}{16 \pi } \int d^4x \sqrt{-g(x)}\, R(x),
\end{eqnarray}
and the matter part corresponds to the black hole core, resulting from stringy effects, and is given by
\begin{eqnarray}
     S_{\rm bare}=\frac{1}{16 \pi } \int d^4x \sqrt{-g(x)}\, \mathcal{L}_{\rm bare}(x),
\end{eqnarray}
where $\mathcal{L}_{\rm bare}(x)$ will be defined later. 
Furthermore, we add a nonlocal term due to the gravitational self-energy interaction term (GSE) given by
\begin{equation}
    S_{\rm GSE }= \frac{1}{16 \pi }  \int d^4x \sqrt{-g(x)} \, \mathcal{L}^{\text{GSE}}(x)
\end{equation}
and we have defined, 
\begin{eqnarray}
  \mathcal{L}^{\text{GSE}}(x):= \int d^4y \sqrt{-g(y)}\, \mathcal{K}(x-y) \mathcal{L}_{\rm bare}(y)\label{ker}
\end{eqnarray}
such that $\mathcal{L}^{\text{GSE}}$ is fixed during the variation  (it depends on $g_{\mu \nu}(y)$), on the other hand, $\mathcal{K}(x-y)$ is the nonlocal kernel that encodes the influence of curvature at point $y$ at point $x$. Note that we can choose 
\begin{equation}
\mathcal{K}(x-y)= \delta(x^0-y^0) \mathcal{R} (\textbf{x}-\textbf{y}),
\end{equation}
where $\mathcal{R} $ is the spatial kernel and it is worth pointing out that this expression is local in time and non-local in space. We emphasize that the superscript ``$0$'' appearing in Eqs.~(2) and (7) denotes evaluation in the static limit (i.e., $k^0=0$ or $x^0=y^0$), whereas the subscript ``$0$'' in Eqs.~(1) and (2) refers to intrinsic parameters of the model, such as the zero-point length $l_0$ or the mass parameter $m_0$. In addition, we have
\begin{equation}
\mathcal{R} (\textbf{x}-\textbf{y}) = \frac{\mathcal{C} (\textbf{x})}{\sqrt{|\textbf{x}-\textbf{y}|^2+l_0^2}} ~,
\end{equation}
where $\mathcal{C} (\textbf{x})$ is some parameter with appropriate units. We can set $\mathcal{C} (\textbf{x})=\rho^{\rm bare} (\textbf{x})$. In many cases, by neglecting the dimensions, one can write $\rho^{\rm bare}=M \delta(\textbf{x})$.
The total action therefore reads 
\begin{eqnarray}
    S= S_{\rm EH}+ S_{\rm GSE }+S_{\rm bare}.
\end{eqnarray}

By performing a variation of the action with respect to the metric $g^{\mu \nu}(x)$, we get
\begin{eqnarray}\notag
  &&  \frac{\delta S}{\delta g^{\mu \nu}(x)} = \int d^4x 
    \frac{\delta \left(\sqrt{-g(x)} R(x)\right)}{\delta g^{\mu \nu}(x)} \\\notag
   &+& \int d^4x 
    \frac{\delta \left( \sqrt{-g(x)} \,[\mathcal{L}_{\rm bare}(x)+\mathcal{L}^{\text{GSE}}(x)]\right)}{\delta g^{\mu \nu}}=0.
\end{eqnarray}

Using $R(x)=g_{\mu \nu}(x) R^{\mu \nu}(x)$ and the relation for the variation of the determinant
\begin{eqnarray}
    \delta \sqrt{-g(x)}=-\frac{1}{2} \sqrt{-g(x)}\,\, g_{\mu \nu}(x) \delta g^{\mu \nu}(x),
\end{eqnarray}
along with the stress-energy tensor defined as 
\begin{eqnarray}
   T^{\rm bare}_{\mu \nu}(x)=-\frac{2}{\sqrt{-g(x)}} \frac{\delta \left(  \sqrt{-g(x)} \, \mathcal{L}_{\rm bare}(x)\right)}{\delta g^{\mu \nu}(x)} ,
\end{eqnarray}
and 
\begin{equation}
   T^{\rm GSE}_{\mu \nu}(x)=-\frac{2}{\sqrt{-g(x)}} \frac{\delta \left(\sqrt{-g(x)} \,\mathcal{L}^{\text{GSE}} (x)\right)}{\delta g^{\mu \nu}(x)},
\end{equation}
we obtain the Einstein field equations, 
\begin{eqnarray}
\label{einstein-field}
 R_{\mu \nu} + \frac{1}{2} g_{\mu \nu} R= 8 \pi \left(T^{\rm bare}_{\mu \nu}+T_{\mu \nu}^{\rm GSE}\right) \equiv 8 \pi \mathcal{T}_{\mu \nu}.
\end{eqnarray}
We have defined the total energy-momentum tensor, $\mathcal{T}_{\mu \nu}$, as the sum of $T^{\rm bare}_{\mu \nu}$ and $T_{\mu \nu}^{\rm GSE}$.
Note that $T_{\mu \nu}^{\rm GSE}$ describes the gravitational self-energy contribution and has the form
${T^{\mu}}_\nu^{\rm GSE}=\left(-\rho^{\rm GSE}, P_r^{\rm GSE}, P_T^{\rm GSE}, P_T^{\rm GSE} \right)$.
In the next Section we will look at the Newtonian limit of $T_{\mu \nu}^{\rm GSE}$, and find that it has a form similar to the electromagnetic energy-momentum tensor under T-duality as obtained in \cite{Gaete:2022ukm} (compare equation (25) of \cite{Gaete:2022ukm} with \eqref{rho-gse} in the next Section).
The total energy-momentum tensor has a form similar to that of $T_{\mu \nu}^{\rm GSE}$, namely ${\mathcal{T}^{\mu}}_{\nu}=\left(-\rho, \mathcal{P}_r,  \mathcal{P}_T,  \mathcal{P}_T \right)$. \\
\\
\textbf{\textit{Newtonian limit for the gravitational self-energy.}}
One particular aspect that we point out is that adding the effect of the gravitational field into the metric in GR does not make sense because GR is a local theory. However, in the present paper the extra term ({\it i.e.} the self-energy due to the gravitational field) emerges as a non-local effect. This gives a new term in the field equations. An example of non-local gravity can be found in \cite{Hehl:2008eu} in the framework of the translational gauge theory of gravity. Here we also have a modified, non-local theory of gravity. Let us estimate the self-energy of this non-local effect. From Eq. \eqref{ker} we can get the energy density via the non-local term 
 \begin{align}
\rho^{\text{GSE}}(t,\textbf{x}) =\int  \mathcal{R} (\textbf{x} -\textbf{y} ) \rho^{\text{bare}}(t,\textbf{y} ) d^3 \textbf{y},
\end{align}
 where we have used $\mathcal{L}_{\rm bare}=-\rho^{\rm bare}$. Under this choice and since we are interested in the static case, there are no dynamical new equations for the matter. 
We can compute the average self-energy using the fact that 
 \begin{equation}
 \left\langle E^{\text{GSE}} \right\rangle :=\frac{1}{2}\int \rho^{\text{GSE}}(\textbf{x})\, d^3 \textbf{x},
 \end{equation}
which gives
\begin{equation}
 \left\langle E^{\text{GSE}} \right\rangle = \frac{1}{2} \int \int \, \frac{\rho^{\text{bare}}(\textbf{x})\rho^{\text{bare}}(\textbf{y})}{\sqrt{|\textbf{x}-\textbf{y}|^2+l_0^2}}\, d^3 \textbf{x}\, d^3 \textbf{y}.
 \end{equation}
  Using the relation for the gravitational potential
 \begin{equation}
V_G(\textbf{x}) = - \int \, \frac{\rho^{\text{bare}}(\textbf{y})}{\sqrt{|\textbf{x}-\textbf{y}|^2+l_0^2}}\, d^3 \textbf{y}~,
 \end{equation}
it follows that
 \begin{equation}
 \left\langle E^{\text{GSE}} \right\rangle = -\frac{1}{2} \int V_G(\textbf{x})\, \rho^{\text{bare}}(\textbf{x})\, d^3 \textbf{x}.
 \end{equation}
Assuming spherical symmetry to simplify the calculations $\textbf{x}\equiv \textbf{r}$, we can use Gauss’s law for gravity,
\begin{equation}
\nabla \cdot \mathbf{g} = -4\pi \rho^{\text{bare}},
\end{equation}
to obtain
\begin{equation}
\langle E^{\text{GSE}} \rangle = \frac{1}{8\pi} \int V_G(r)\, (\nabla \cdot \mathbf{g})\, d^3\mathbf{r}. 
\end{equation}
By making use of the identity
\begin{equation}
\nabla \cdot (V_G\, \mathbf{g}) = V_G\, (\nabla \cdot \mathbf{g}) + \mathbf{g} \cdot (\nabla V_G) 
\end{equation}
then using the divergence theorem 
\begin{equation}
\int_V \nabla \cdot (V_G\, \mathbf{g})\, d^3\mathbf{r} = \oint_{\partial V} V_G\, \mathbf{g} \cdot d\mathbf{A} \to 0.
\end{equation}
it follows
\begin{equation}
\langle E^{\text{GSE}} \rangle =
- \frac{1}{8\pi} \int \mathbf{g} \cdot (\nabla V_G)\, d^3\mathbf{r}. 
\end{equation}
For the distant observer, only this term gives a contribution. To obtain the gravitational self-energy term to the energy-momentum tensor we will use a coordinate-independent quantity observable by an asymptotic observer in the Newtonian limit, namely the Newtonian gravitation field defined as 
\begin{eqnarray}
    \textbf{g}=-\nabla V_G(r)= -\frac{Mr}{(r^2 + l_0^2)^{3/2}}\hat{r}.
\end{eqnarray}
The gravitational self-energy term is then proportional to the volume integral of the square of $\textbf{g}$
\begin{eqnarray}\label{GSE}
\langle E^{\rm GSE} \rangle &=& \frac{1}{8\pi}\int_0^r \textbf{g}^2(r') d^3\textbf{r}' \nonumber \\
&=&-\frac{5 M^2 r^3}{16(r^2+l_0^2)^2}-\frac{3 M^2 l_0^2 r}{16(r^2+l_0^2)^2}\\
&+&\frac{3  M^2}{16 l_0} \arctan(\frac{r}{l_0}). \nonumber
\end{eqnarray}
The last equation will play an important role in our discussion of the black hole solution and the emergence of dark matter.

\section{Regular black hole solutions}
We saw that the Einstein field equations are modified by the extra GSE term, $\mathcal{T}_{\mu\nu}^{\mathrm{GSE}}$, on the right-hand side of Eq.~\eqref{einstein-field}. Consequently, as we shall see, this leads to a new and more general black hole solution compared to the one presented in Ref. ~\cite{Nicolini:2019irw}.

 To obtain the spacetime geometry from Eq. \eqref{einstein-field}, or equivalently written in terms of Einstein tensor as $G_{\mu \nu}=8 \pi  \mathcal{T}_{\mu \nu}$,  we first assume a static and spherically spacetime of the form
\begin{equation}
    ds^2=g_{tt}dt^2+g_{rr}dr^2+r^2(d\theta^2+\sin^2\theta d\phi^2). \label{metric}
\end{equation}

By imposing the standard choice,
\begin{equation}
   -g_{tt}=g_{rr}^{-1}=f(r),
\end{equation}
and by means of the Einstein field equations \eqref{einstein-field} we get for the Einstein tensor components ${G^t}_t={G^r}_r$. 
This results in an equivalent equation of state ${\mathcal{T}^{t}}_{t}={\mathcal{T}^{r}}_{r}$, or equivalently, $\rho=-\mathcal{P}_r$. Next, from the conservation of the energy-momentum tensor, 
\begin{eqnarray}
    \nabla_\mu  \mathcal{T}^{\mu \nu}=0,
\end{eqnarray}
one can obtain 
\begin{eqnarray}
    \frac{d \mathcal{P}_r}{dr}=-\frac{1}{2 g_{tt}}\frac{dg_{tt}}{dr}(\rho+\mathcal{P}_r)+\frac{2}{r}(\mathcal{P}_T-\mathcal{P}_r).
\end{eqnarray}

From the last equation and using the condition $\rho=-\mathcal{P}_r$, it follows that
\begin{eqnarray}
   \mathcal{P}_T=-\rho-\frac{r}{2} \frac{d \rho}{dr}.
\end{eqnarray}

In general, one can write for the total energy-momentum tensor the following relation (similar to \cite{Gaete:2022ukm})
\begin{eqnarray}
    \mathcal{T}_{\mu \nu}=-\rho  g_{\mu \nu}+\tau_{\mu \nu} \ ,
\end{eqnarray}
where $\rho$ is now the total energy density given by 
\begin{eqnarray}
    \rho=\rho^{\rm bare}+\rho^{\rm GSE},\,\,\, \text{and}\,\,\,\,\tau_{\mu \nu}&=& \tau_{\mu \nu}^{\rm bare}+ \tau_{\mu \nu}^{\rm GSE}
\end{eqnarray}
in which
\begin{eqnarray}
    \tau_{\mu \nu}&=&{\rm diag} \left(0, 0, -(r/2)\, (\partial_r \rho), -(r/2)\, (\partial_r \rho)  \right).
\end{eqnarray} 

Using Poisson's equation gives an energy density function for the bare matter as
\cite{Nicolini:2019irw}
\begin{equation}
\rho^{\rm bare}(r)= \frac{1}{4\pi}\nabla^2 V_G(r)=\frac{3 l_0^2 M}{4 \pi \left( r^2+l_0^2\right)^{5/2}}.\label{density}
\end{equation}

Using the energy density in Eq. \eqref{density} we obtained the radial, $P_r$, and transverse,  $P_T$, pressure given as
\begin{eqnarray}
    P_r^{\rm bare}&=-&\rho^{\rm bare}
    =-\frac{3 l_0^2 M}{4 \pi \left( r^2+l_0^2\right)^{5/2}},\\
    P_T^{\rm bare}&=&-\rho^{\rm bare}-\frac{r}{2} \left(\partial_r \rho^{\rm bare}\right) \\
    &=&\frac{3 l_0^2 M (3r^2-2 l_0^2)}{8 \pi \left( r^2+l_0^2\right)^{7/2}} ~.\nonumber
\end{eqnarray}

Further, by using the bare matter energy-momentum tensor it can be seen that the Strong
Energy Condition ({\it i.e.} $\rho^{\rm bare}+\sum_i P^{\rm bare}_i \geq 0$) is violated in the region $r < \sqrt{2/3} \,l_0$. This shows the non-classical nature of spacetime under T-duality effects. As shown in \cite{Jusufi:2024dtr}, such violations play an important role in evading the black hole singularity theorem. 

To compute the average energy density for the self-interaction energy, we use Eq. \eqref{GSE} and the following relation
\begin{eqnarray}
\label{rho-gse}
\rho^{\rm GSE}  = \frac{1}{4\pi r^2}\frac{\partial}{\partial r}(\langle E^{\rm GSE} \rangle) = \frac{M^2r^2}{8\pi(r^2 + l_0^2)^3}.
\end{eqnarray}

Using \eqref{rho-gse} one obtain the pressure components of the GSE energy-momentum tensor as
\begin{equation}\label{Pressurer}
   P^{\rm GSE}_r=-\rho^{\rm GSE}
    =-\frac{M^2r^2}{8\pi(r^2 + l_0^2)^3}.
\end{equation}
and
\begin{eqnarray}
    P_T^{\rm GSE}&=&-\rho^{\rm GSE}-\frac{r}{2} \left(\partial_r \rho^{\rm GSE}\right) \nonumber \\
    &=&\frac{r^2 M^2 (r^2-2 l_0^2)}{8 \pi \left( r^2+l_0^2\right)^4}.
\end{eqnarray}

Using the GSE energy-momentum tensor, one can see that the Strong Energy Condition ({\it i.e.} $\rho^{\rm GSE}+\sum_i P^{\rm GSE}_i \geq 0$) is again violated in the region $r < \sqrt{2} \,l_0$. This again highlights that under T-duality, spacetimes can behave non-classically.

Furthermore, we can assume that the total mass of the system is given by the bare mass plus the mass/energy stored in the gravitational field
\begin{eqnarray}
     m(r)=m^{\rm bare}(r)+ m^{\rm GSE}(r).
\end{eqnarray}
Specifically, we can compute the mass enclosed in some region by using the energy density for the bare matter and non-local gravitational energy 
\begin{eqnarray}
    m(r)=4 \pi \int_0^r \left[\rho^{\rm bare}(r') +\rho^{\rm GSE}(r') \right]r'^2 dr'.
\end{eqnarray}

In this way, by solving the above integral for the metric function, we get the following result 
\begin{eqnarray}
    f(r)=1-\frac{2 m(r)}{r},
\end{eqnarray}
yielding 
\begin{eqnarray}\notag
    f(r)&=&1-\frac{2 M r^2}{ (r^2+l_0^2)^{3/2}}+\frac{5 M^2 r^2}{8(r^2+l_0^2)^2}+\frac{3 M^2 l_0^2 }{8(r^2+l_0^2)^2}\\
    &-&\frac{3 M^2}{ 8 l_0 r} \arctan(\frac{r}{l_0}).
\end{eqnarray}

The above solution can be further written in a more compact form as
\begin{eqnarray}\label{metric1}
    f(r)=1-\frac{2 M r^2}{ (r^2+l_0^2)^{3/2}}+\frac{ M^2 r^2  }{(r^2+l_0^2)^2} F(r)\label{metric1},
\end{eqnarray}
where 
\begin{equation}
    F(r)=\frac{5}{8}+\frac{3l_0^2}{8 r^2}-\frac{3 (r^2+l_0^2)^2}{8 l_0 r^3}\arctan(\frac{r}{l_0}).
\end{equation}

There are several properties of the solution given by Eq.~\eqref{metric1}, which are listed below:\\
\begin{enumerate}[label=(\roman*)]
    \item Firstly, it describes a regular and singularity-free geometry representing a completely neutral object, which can be either a particle or a black hole depending on the mass parameter. 

\item Secondly, the solution possess more ADM mass than would be predicted by classical considerations (GR). The extra corrections to the mass emerge due to the non-local nature of the gravitational self-energy term.

\item Thirdly, in the limit $F(r)\to 1$ it remarkably reproduces the regular Ayon-Beato–Garcia-type solution \cite{Ayon-Beato:1998hmi}, where the role of the electric charge $q$ is played by the zero-point length $l_0$, and the black hole charge $Q$ is effectively replaced by the bare mass $M$. 

\item Fourthly, the solution $f(r)$ is an even function of the radius coordinate $r$, which is a good feature for the smoothness of the metric (see \cite{ZhouModesto2023,Giacchini:2021pmr} and recently \cite{Antonelli:2025zxh}). The good parity of our metric is an important advantage compared to other regular black hole models, such as the famous Hayward metric. 
\item Finally, as we will show later, the solution leads to a stable, cold remnant in the final stages of evaporation, with a mass on the order of the Planck mass. Since the solution does not involve an electric charge, this result is expected, namely, a stable remnant with no further decay via Schwinger pair production. \cite{Gibbons:1975kk,Page:2006cm,Nicolini:2017hnu}.
\end{enumerate}
 \begin{figure}[ht!]
		\centering
	\includegraphics[scale=0.60]{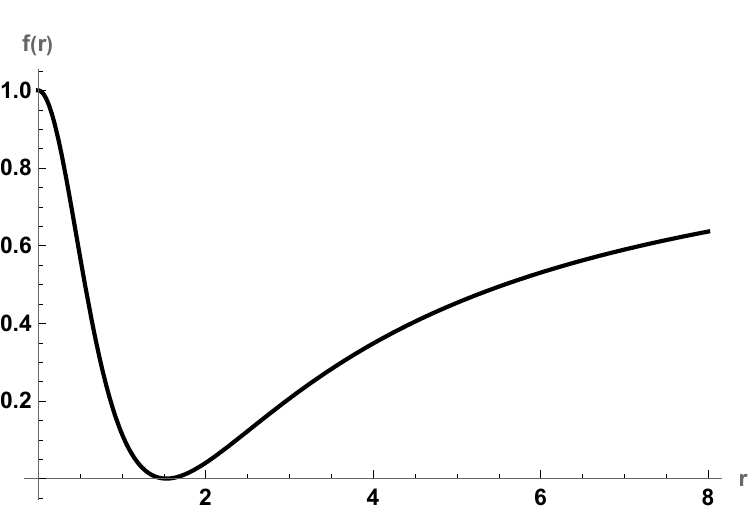}
\caption{The plot shows the extremal configuration for the black hole using the metric function \eqref{metric1} with $r_{\rm ext}=1.53231$ and $M_{\rm ext}=1.16537$. The Planck length is set to unity, i.e. $l_0=l_{Pl}=1$.  }
	\end{figure}
\textit{\textbf{Discussion on the ADM mass.}} Let us elaborate on the ADM mass or the mass measured by an observer located at some asymptotic region. In the region $r \gg l_0$, from the metric function we get the following behavior 
\begin{eqnarray}
    f(r)=1-\frac{2 \left(M+\frac{3 \pi M^2}{32 l_0} \right)}{ r}+\dots,
\end{eqnarray}
we thus find the ADM mass $\mathcal{M}$ of the black hole is corrected due to the gravitational field and is given by 
\begin{eqnarray}
   \mathcal{M}=M\left(1+\frac{3 \pi M}{32 l_0}\right).\label{ADM}
\end{eqnarray}
By making use of the relations 
\begin{eqnarray}
    l_{Pl}=\sqrt{\frac{\hbar G}{c^3}},\,\,\,   M_{Pl}=\sqrt{\frac{\hbar c}{G}},
\end{eqnarray}
and assuming $l_0 \sim  l_{Pl}$, we can write the ADM mass as follows
\begin{eqnarray}\label{mass}
   \mathcal{M} = M \left(1+ \frac{3 \pi M}{32 M_{Pl}} \right).
\end{eqnarray}

The second term in Eq. \eqref{mass} or Eq. \eqref{ADM} represents the energy contribution from the regularized gravitational self-energy.
    
Such an effect on the spacetime geometry was not included in the original derivation reported in \cite{Nicolini:2019irw}. 
Interestingly, there is a striking similarity between the gravitational field effect and the electric field effect discussed in \cite{Gaete:2022ukm}. In fact, the corrections due to the gravitational field coincide with the results in \cite{Jusufi:2023tdn}, where the corrected energy was derived from the Schr\"odinger-Newton equation. As we shall see in more detail later, for small masses, the corrections due to self-gravitational effects are perfectly reasonable. However, for very large masses, such as those of supermassive black holes, the corrections become extraordinarily large, which is not physically reasonable.

To address this issue with the large mass limit we use a T-duality inspired approach where there is a duality between $M$ and $M^{-1}$ i.e. there is a duality between the large mass and small mass limits. For example \cite{Carr:2015nqa} studied the particle-black hole correspondence and the role of the duality of mass $M \leftrightarrow  M^{-1} $ which implies the Generalized Uncertainty Principle.  It was shown that depending on the mass value one has classical, quantum, relativistic and quantum gravity domains. Our ADM mass equation only yields physically acceptable solutions in one domain. Therefore, we need to extend the equation so that it remains valid across the entire domain. To do this, let us define a function that is invariant under $M  \to M_{Pl}^2/M$, is bounded and reduces to $ M_{Pl}/M$, in the large mass limit.

The proposed mass inversion is motivated by the standard particle--black-hole correspondence:
the Schwarzschild radius $r_s \sim GM$ increases with $M$, whereas the Compton wavelength
$\lambda_C \sim 1/M$ decreases with $M$, and the two meet at the Planck scale $M\sim M_{\rm Pl}$.
This motivates identifying a ``dual'' description under $M \to M_{\rm Pl}^2/M$, which exchanges the
large-mass (classical black hole) and small-mass (particle/quantum) regimes and is closely tied to
GUP/T-duality reasoning (see Ref.~\cite{Carr:2015nqa} and related discussions in
Refs.~\cite{Padmanabhan:1996ap,Jusufi:2024utf}). Below we introduce an interpolation function $\zeta(x)$ to be bounded and
invariant under $x \to 1/x$, so that the ADM correction saturates in both asymptotic regimes while
preserving the desired Planck-scale behavior. 

We propose an interpolation function given by
\begin{equation}
\zeta(x)=\frac{2x}{1+x^2},\label{intfunction}
\end{equation}
where $x=M/M_{Pl}$ which under the T-duality transformation of the mass $M  \to M_{Pl}^2/M$, yielding $x  \to 1/x$, and hence is invariant $\zeta(x)=\zeta(1/x)$.

The modification $M/M_{\rm Pl} \rightarrow \zeta(M/M_{\rm Pl})$ can be motivated from several complementary theoretical considerations. First, in T-duality inspired frameworks, physical quantities are expected to be invariant under the transformation $M \rightarrow M_{\rm Pl}^2/M$, reflecting the symmetry between large and small mass (or equivalently, length) scales. This duality is closely related to the particle–black hole correspondence, where the Schwarzschild radius $r_s \sim GM$ and the Compton wavelength $\lambda_C \sim 1/M$ intersect at the Planck scale.

Second, the original ADM correction term grows unbounded for large masses, which is physically problematic in the astrophysical regime. Therefore, it is natural to introduce a bounded interpolation function that regularizes this behavior while preserving the correct limits in both $M \ll M_{\rm Pl}$ and $M \gg M_{\rm Pl}$ regimes.

From an effective field theory perspective, this modification can be interpreted as a phenomenological parametrization of higher-order, non-local corrections that become relevant near the Planck scale, while remaining negligible in the classical regime.

The complete ADM mass is now proposed as follows
\begin{eqnarray}
   \mathcal{M} = M \left(1+ \frac{3 \pi }{32}  \zeta(M/M_{Pl}) \right),
\end{eqnarray}
or equivalently,
\begin{eqnarray}
   \mathcal{M} = M \left(1+ \frac{6 \pi (M/M_{Pl})}{32 \left(1+(M/M_{Pl})^2\right)} \right).\label{ADM1}
\end{eqnarray}
By means of the invariance of the function $\zeta(M/M_{Pl})=\zeta(M_{Pl}/M)$, under the T-duality transformation of mass $M  \to M_{Pl}^2/M$,  we can also write the ADM mass as 
\begin{eqnarray}
   \mathcal{M} = M \left(1+ \frac{6 \pi (M_{Pl}/M)}{32 \left(1+(M_{Pl}/M)^2\right)} \right).\label{ADM11}
\end{eqnarray}
Depending on the mass value, there are three interesting domains that can be elaborated and the last two relations should give equivalent results.

 \begin{enumerate}[label=(\roman*)]
\item {\bf Particle sector}: The first case corresponds to the region $M\ll M_{Pl}$. This represents the \emph{particle sector} and from the ADM mass \eqref{ADM1} or \eqref{ADM11} we find that 
\begin{eqnarray}
  \mathcal{M} = M \left(1+ \frac{3\pi M}{16 M_{Pl}} \right),
\end{eqnarray}
which basically means that $\mathcal{M} \simeq M$  in the limit $M\ll M_{Pl}$. Such a relation is expected and is consistent with observations, since the ADM mass of elementary particles is predominantly determined by their bare mass.

\item {\bf Particle-Black Hole sector}: In the case $M=M_{Pl}$, from the ADM mass \eqref{ADM1} or \eqref{ADM11} we obtain the form of Eq. \eqref{mass} which reads,
\begin{eqnarray}
\label{bh-part}
  \mathcal{M} = M \left(1+ \frac{3 \pi M }{32 M_{Pl} } \right)=1.2943\, M_{Pl}.
\end{eqnarray}
We shall refer to this as the \emph{particle–black hole} sector, which is a particularly interesting feature of our solution and implies the existence of an event horizon and therefore a black hole is indeed possible. We refer to this as the \emph{particle–black hole} case, which represents the most physically compelling situation. However, it is interesting to see that the solution has more ADM mass than would be predicted by classical GR. The extra corrections to the mass emerge because of the non-local nature of gravity. This will lead to phenomenological implications related to the problem of dark matter, which will be addressed in the last section. 

\item {\bf Astrophysical Black Hole sector}: Finally, in the region $M\gg M_{Pl}$, or the large mass limit, and by means of \eqref{ADM1} or \eqref{ADM11}, we get for the ADM mass 
\begin{eqnarray}
   \mathcal{M} = M \left(1+ \frac{3 \pi M_{Pl}}{16 M} \right).
\end{eqnarray}
As expected in the limit $M\gg M_{Pl}$, the total mass is dominated by the bare mass and the ADM mass reduces to the bare mass, i.e., $\mathcal{M} \simeq M$. This is the region of supermassive black holes where the corrections to the mass are saturated to a constant $M_{Pl}$ term and the apparent divergence is removed. This confirms that we do obtain physically acceptable results for the ADM mass in both regions.
\end{enumerate}

It is interesting to point out that once we introduce the interpolation function given by Eq. \eqref{intfunction} the fundamental length $l_0$ becomes a function of $M$. As we will see from the discussion below $l_0 (M)$ varies between $l_{Pl}$ (when $M$ is the Planck mass) and $l_{Pl}/2$ at the two extremes of small mass and large mass. Specifically, by means of Eq. \eqref{ADM} and by using Eq. \eqref{ADM1} and \eqref{ADM11}, we will have the following effective length relation
\begin{equation}
 l_0 (M)  \sim  \frac{l_{Pl}}{2} \left[1+ \left(\frac{M}{M_{Pl}}\right)^2 \right],\,\,  \text{when}\,\,\,M\ll M_{Pl}\label{funlength0},
\end{equation}
and 
\begin{equation}
 l_0 (M)  \sim  \frac{l_{Pl}}{2} \left[1+ \left(\frac{M_{Pl}}{M}\right)^2 \right],\,\,  \text{when}\,\,\,M\gg M_{Pl}. \label{funlength00}
\end{equation}
At the special value $M=M_{Pl}$, we obtain $l_0 \sim  l_{Pl}$, as expected. \\

\textit{\textbf{Black hole remnants.}}
The existence of extremal configurations with extremal mass $M_{\rm ext} \sim M_{Pl}$, (the particle-black hole sector in point (ii) above) characterized by a single degenerate horizon $r_{\rm ext}=r_-=r_+$, can be studied numerically by solving the following conditions:
\begin{equation}
f(r_{\rm ext},M_{\rm ext})=0,\qquad \left. \frac{\partial }{\partial r} f(r,M) \right|_{r = r_{\rm ext},M=M_{\rm ext}} = 0.\label{conditions}
\end{equation}
Using the metric \eqref{metric1} and working in Planck units we obtain 
\begin{equation}
r_{\rm ext}=1.53231 [l_{\text{Pl}} ],\qquad  M_{\rm ext}=1.16537  [M_{\text{Pl}}].
\end{equation}
For the total ADM mass using \eqref{ADM1} we obtain
\begin{equation}
\mathcal{M}_{\rm ext} = 1.50462[M_{\text{Pl}}].\label{extrimalm}
\end{equation}

In Fig. 1 we have shown the plot of the metric function \eqref{metric1} for the extremal black hole configuration with a single degenerate horizon. It should be noted that $r_{\rm ext}$  and $\mathcal{M}_{\rm ext}$ are greater than the extremal radius obtained without including gravitational self-energy corrections, which is given by $r_{\rm ext} = \sqrt{2} \,\,[l_{\text{Pl}}]$ ~\cite{Nicolini:2019irw}. Similarly, the extremal mass is also larger than in the uncorrected case, where it is given by $M_{\rm ext} = 3\sqrt{3}/4 \,\, [M_{\text{Pl}}]$ (see also~\cite{Nicolini:2019irw}). 

We note that the extremal remnant configuration discussed in this subsection corresponds to a zero-temperature state where the Hawking evaporation process effectively halts. Within the present static and semi-classical framework, this extremal solution appears as a natural endpoint of evaporation.

However, the uniqueness of this remnant depends on the dynamical details of the evaporation process, including possible backreaction effects and quantum gravity corrections beyond the semi-classical approximation. In particular, different initial conditions or alternative non-local completions of the theory could, in principle, lead to distinct remnant configurations. Therefore, while our model predicts a stable endpoint, a rigorous proof of uniqueness would require a fully dynamical treatment of black hole evaporation in the underlying non-local theory.\\

\textit{\textbf{Thermodynamics and stability.}}
Here we show that the particle-black hole sector solutions of point (ii) above are thermodynamically stable objects with a vanishing Hawking temperature and positive heat capacity. The Hawking temperature can be obtained using the metric function, $f(r)$, from \eqref{metric1} 
\begin{eqnarray}
\label{temp1}
T= \left.\frac{f'(r)}{4 \pi}\right|_{r_+}.
\end{eqnarray}

Using the fact that $f(r_+)=0$, we can calculate the mass parameter $M$, which reads
\begin{eqnarray}\notag
\label{massr}
M(r_+)&=&\frac{(l_0^2+r^2_+)^{1/2}\sqrt{r_+\left(r_+-F(r_+)(l_0^2+r^2_+)\right)}}{F(r_+) r_+^2} \\&+& \frac{ (l_0^2+r^2_+)^{1/2}}{F(r_+)r_+}.
\end{eqnarray}

 Finally we can use \eqref{temp1} and \eqref{massr} to calculate the heat capacity giving
\begin{eqnarray}
C = \left. \frac{\partial M}{\partial T} \right|_{r_+}=  \frac{\partial M}{\partial r_+}  \left(\frac{\partial T}{\partial r_+} \right)^{-1}.
\end{eqnarray}

From Fig.~2 one can see that the Hawking temperature reaches a peak and then vanishes at the extremal horizon radius.  While the temperature of a singular Schwarzschild black holes increases without bound as the mass decreases, for regular black holes the temperature initially increases, reaches a maximum and then goes to zero. The result of this is that a black hole will initially radiate according the the usual Hawking temperature, but as its mass decreases the evaporation will slow down and eventually go to zero as the temperature goes to zero, leaving a non-radiating, cold remnant with a mass around the Planck mass.

One can see from Fig. 2 that there exists a region with negative Hawking temperature, but such an effect cannot be probed by an outside observer, so it is not of physical interest. However, the deeper explanation lies in the violation of the Strong Energy Condition (SEC) in the interior region given by $r < \sqrt{2/3}\,l_0 $, which has a de Sitter core. Specifically in that region, quantum gravity essentially acts as antigravity, and this mechanism is necessary to evade the singularity theorem (see, for example, \cite{Jusufi:2024dtr}).

Furthermore, we can see from Fig. 3 that the system undergoes a phase transition at the point of maximum temperature. The heat capacity changes sign at this transition, indicating a shift from thermodynamically unstable (negative heat capacity) for large black holes to stable black hole configurations (positive heat capacity) as it approaches the extremal configuration. These objects may have been formed either in the early universe from density perturbations or as remnants of evaporating massive black holes via Hawking radiation.

In addition, in some of the above papers, the Bardeen spacetime was used; unfortunately, such a solution requires the existence of magnetic monopoles, for which there is currently no experimental evidence. In contrast, a T-duality-based solution such as the solution presented in this paper is purely neutral, with the role of the magnetic charge effectively being played by the zero-point length. Our solution does not require the existence of magnetic monopoles.
An important point that follows from this property is that neither the solution obtained in \cite{Nicolini:2019irw} nor our metric \eqref{metric1} suffers from Schwinger instability. 

\begin{figure}[ht!]
		\centering
	\includegraphics[scale=0.65]{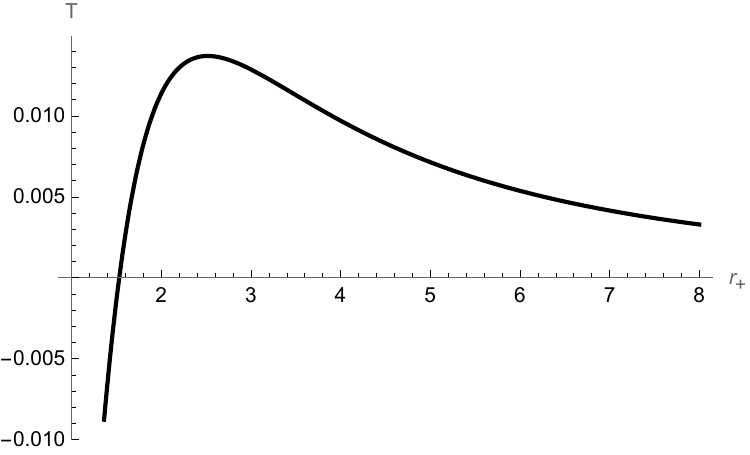}
\caption{The plot shows Hawking temperature for the extremal configuration. }
	\end{figure}

    \begin{figure}[ht!]
		\centering
	\includegraphics[scale=0.65]{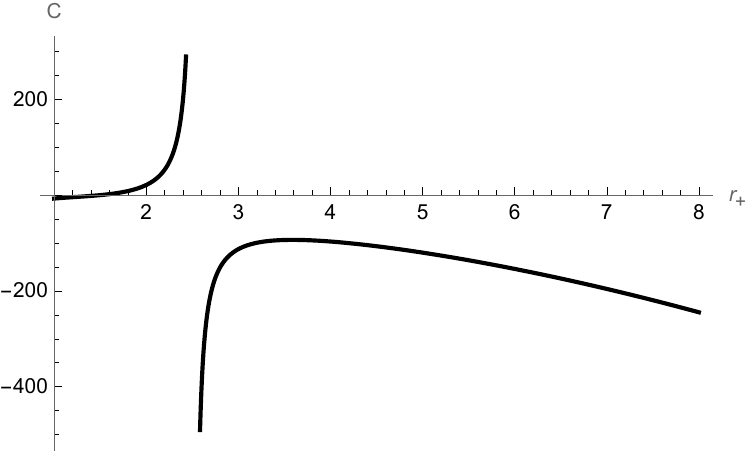}
\caption{The plot shows the heat capacity for the extremal configuration.}
	\end{figure}

\section{Possible phenomenological implications: dark matter from black hole remnants}

 Currently the form of the missing mass, or dark matter, is still and open, unsolved puzzle. The existence of dark matter is supported by multiple observations: gravitational lensing, cosmic microwave background measurements, and large-scale structure formation.  Dark matter is estimated to account for about 25\% of the energy density of the Universe.  Despite extensive searches for potential dark matter candidates like WIMPS \cite{LeeWeinberg1977,Goldberg1983PRL,Ellis1984NuPhB238} and axions \cite{PecceiQuinn1977PRL,Preskill:1982cy,Abbott:1982af}, no dark matter particle has been detected. 
In this section we  elaborate more on the idea that the particle-black hole sector from the previous section ({\it i.e.} option (ii) given by \eqref{bh-part})  could play the role of dark matter particles. The possibility that dark matter comes from primordial black holes (PBHs) has been proposed in many studies ~\cite{Carr:2015nqa, PBH1, PBH2, PBH3,PBH4,Mazde:2022sdx,Carr:2022ndy}. Observational limits and constraints on PBH as dark matter have been obtained by using  evaporation via Hawking radiation, gravitational lensing, dynamical interactions, accretion, and gravitational-wave emission (see Fig. 4 and \cite{Carr:2025kdk} for more information). 
Using the usual singular PBHs as dark matter faces  serious issues. For instance, conventional calculations~\cite{Hawking:1974rv} predict that a black hole has a Hawking temperature $T \propto 1/M$ and a luminosity $P \propto 1/M^2$, which implies that a black hole with mass $10^{11}$~kg would evaporate in a time frame of approximately $2.6 \times 10^{9}$ years and the final stages of this evaporation via Hawking radiation would appear as an explosive event leaving behind no remnant. Up to the present, no such explosions have been observed, which then provides strong constraints on low-mass primordial black holes  as dark matter. Here we propose that the non-singular PBHs of the particle-black sector (similar to \cite{Calza:2024fzo,Calza:2024xdh,Calza:2025mwn,Davies:2024ysj,Carr:2025auw,Asmanoglu:2025agc} and also \cite{Davies:2024ysj,Carr:2025auw}) can indeed be good candidates for dark matter. \\ 
\\
\\

\textbf{\textit{Production of primordial black holes}}
Planck mass, non-singular PBH of the type discussed in point (ii) of the last section ({\it i.e.} the particle-black hole sector) can form after inflation, during the radiation-dominated era, and before Big Bang nucleosynthesis. The corresponding primordial density fluctuations  take the form \cite{Carr:1975qj}
\begin{equation}
    \delta = \epsilon \left( \frac{\mathcal{M}}{\mathcal{M}_{\rm int}} \right)^{-n},
\end{equation}
where $\mathcal{M}_{\rm int}$ is the initial mass within the cosmological horizon at the moment of formation, $0 \leq \epsilon \leq 1$ and $n \geq 0$.

Under this assumption that these Planck mass, non-singular PBH remnants form shortly after the end of inflation, then the  relation for the initial mass density of primordial black holes is \cite{Davies:2024ysj}
\begin{equation}
    \rho^{\text{initial}}_{\text{PBH}} = \mathcal{A} \left(\sqrt{\frac{\mathcal{M}_{\rm int}}{\mathcal{M}_{\text{lower}}}} -\sqrt{\frac{\mathcal{M}_{\rm int}}{\mathcal{M}_{\text{upper}}}}\right),
\end{equation}
where we have defined 
\begin{equation}
\mathcal{A}=2\mu_{\rm int} F \epsilon \exp \left( -\frac{\beta^4}{2\epsilon^2} \right),
\end{equation}
where $\mu_{\rm int}$ is the density of the Universe at the time  the PBHs form, $F$ is the ratio of the number density today to number density initially, and $\beta$ is fractional collapse parameter. In addition, $\mathcal{M}_{\text{upper}}$ and $\mathcal{M}_{\text{lower}}$ are the largest and smallest masses of the primordial black holes formed. For $\mathcal{M}_{\text{upper}} \gg \mathcal{M}_{\text{lower}}$, it follows that \cite{Davies:2024ysj}
\begin{equation}
    \Omega^{\text{today}}_{\text{PBH0}}  \sim \frac{\mathcal{B}}{\sqrt{\mathcal{M}_{\text{lower}}}}.\label{dm0}
\end{equation}
where the subscript $`0'$ indicates the quantities evaluated ``today", namely $z=0$. In addition we have defined $\mathcal{B}=\mathcal{A} \sqrt{\mathcal{M}_{\rm int}}$. 
Based on this, the fraction of dark matter that could be PBH remnants in terms of the dark matter density parameter reads \cite{Davies:2024ysj}
\begin{equation}
\Omega_{\text{PBH}}(z) \sim \Omega^{\text{today}}_{\text{PBH0}} (1+z)^3 .
\end{equation}
    
Specifically, in~\cite{Davies:2024ysj}, it was pointed out that since non-singular black holes do not fully evaporate, it is natural to choose $\mathcal{M}_{\text{lower}} \approx \mathcal{M}_{\text{ext}} \approx M_{Pl}$ and $\mathcal{M}_{\text{upper}} \approx 10^{22}~\text{kg}$. However, since $\mathcal{M}_{\text{upper}} \gg \mathcal{M}_{\text{lower}}$, to calculate the dark matter density parameter we can basically use Eq. \eqref{dm0} along with a specific choice of parameter $\mathcal{B}$. Specifically, using the extremal mass from Eq. \eqref{extrimalm} and by setting $\mathcal{B}=0.000045$ in units kg$^{1/2}$ we obtain reasonable values for the dark matter parameter $\Omega^{\text{today}}_{\text{PBH0}} \approx 0.25$. This value is consistent with observational data and lies within the expected range. 

    \begin{figure}[ht!]
		\centering
	\includegraphics[scale=0.9]{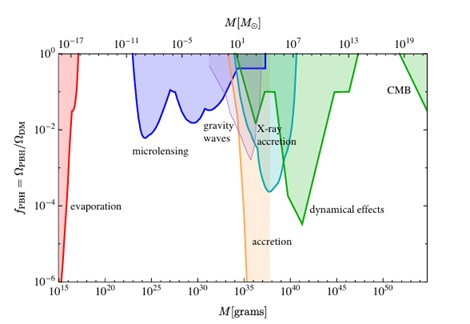}
\caption{The plot shows constraints on the fraction of dark matter as primordial black holes. Note that the shaded regions are excluded under standard assumptions. Specifically, the bounds shown are from evaporation (red), microlensing (blue), gravity-wave (purple), accretion (yellow), X-ray accretion (light blue), and dynamical (green) effects, including the CMB dipole (the plot is taken from \cite{Carr:2025kdk}). Our dark-matter candidate is the \emph{Planck-mass extremal remnant} of the evaporation process, for which $T_H\to 0$.}
	\end{figure}
    
Finally, assuming the current $\rm \Lambda$CDM cosmological model, with matter, radiation, and dark energy (cosmological constant), we can write the following 
\begin{equation}
   \frac{H^2(z)}{H_0^2}=\Omega_{\rm M0}^{\rm \Lambda CDM}(1+z)^3+\Omega_{\rm R0}^{\rm \Lambda CDM}(1+z)^4+\Omega_{\rm \Lambda}^{\rm \Lambda CDM},
\end{equation}
where $\Omega_{\rm M0}$ is the total matter parameter evaluated currently, $\Omega_{\rm R0}$ is the radiation parameter evaluated currently, and similarly $\Omega_{\rm \Lambda}^{\rm \Lambda CDM}$ is the density parameter of dark energy. Furthermore, the total matter parameter will be a sum of the baryonic matter and dark matter parameters given by 
\begin{equation}
  \Omega_{\rm M0}^{\rm \Lambda CDM}=\Omega_{\rm B0}^{\rm \Lambda CDM}+\Omega_{\rm DM0}^{\rm \Lambda CDM},
\end{equation}
where we have defined $\Omega_{\rm DM0}^{\rm \Lambda CDM}=\Omega^{\text{today}}_{\text{PBH0}}.$
This result can explain the observed effect attributed to dark matter using primordial remnants.

\section{Conclusions}
We have presented a novel approach to incorporating gravitational self-energy effects to obtain the spacetime geometry of regular black holes by employing a non-local, T-duality-inspired gravitational theory. By introducing an additional energy-momentum tensor into the Einstein equations, we account for the non-local effect of the self-energy of the gravitational field in a way that is observable by an asymptotic observer in the weak-gravity limit. This framework modifies the ADM mass with a finite, regularized gravitational contribution, resulting in a regular black hole solution of the Ayon-Beato-Garcia type---but in a purely neutral form, without any electric charge. This makes the solution particularly interesting because it does not suffer from Schwinger instability, unlike other regular, electrically charged solutions.

Another interesting and important feature of our solution is that $f(r)$ is an even function of the radial coordinate $r$. This property ensures the smoothness of the metric and provides a significant advantage compared to other regular black hole models, such as the Hayward metric.

One of the most intriguing implications of our analysis is the existence of extremal configurations, or \textit{particle–black hole} objects, with masses of the order of the Planck mass. These objects are shown to be thermodynamically stable, with vanishing Hawking temperature and positive heat capacity. They may be produced either as black hole remnants from the final stages of Hawking evaporation or as primordial black holes formed in the early Universe during the Big Bang. Such stable \textit{particle–black holes} could contribute to the missing mass in the Universe, ({\it i.e.} could serve as dark matter). A central finding of the present work is that gravitational self-energy corrections become significant as the mass approaches the Planck scale. In this regime, the corrections are non-negligible, and it appears that such small objects possess more ADM mass than predicted by classical considerations. We argued that these particle-black hole objects could play the role of dark-matter particles.

Our calculations of the gravitational self-energy are performed in the Newtonian limit, while the full picture remains unknown, particularly the explicit form of the underlying Lagrangian. On the other hand, singularities are expected to arise in the deeply nonlinear regime of general relativity. Therefore, a potential criticism of our approach is that the Newtonian limit breaks down well before one reaches the strong field region near the singularity. We propose that the Newtonian limit modification due to the gravitational self-energy persists into the strong-field regime. 

At present, a first-principles derivation of the gravitational self-energy contribution in the full strong-field regime is not available. Our construction is based on the Newtonian limit, where the notion of gravitational self-energy can be defined in a coordinate-independent manner.

We therefore interpret our results as indicating that the Newtonian self-energy correction captures the infrared limit of a more general non-local gravitational theory. While it is plausible that such non-local effects persist in the strong-field regime—particularly given their origin in zero-point length and T-duality arguments—this remains an assumption of the present model.

A complete validation of this extension would require deriving the full non-local effective action and solving the corresponding field equations beyond the weak-field approximation. We leave this as an important direction for future work.

In the present framework the gravitational self-energy term originates from the same non-local/zero-point-length
structure that regularizes the static potential at short distances. We therefore interpret the Newtonian derivation
as fixing the \emph{infrared limit} of an underlying non-local effective action, and assume that the corresponding
non-local stress-energy contribution remains operative in the non-linear regime. A direct consequence is that the
effective source violates the strong energy condition in a compact region (as already found here), so the focusing
condition in the Raychaudhuri equation can be avoided and curvature singularities need not form. This mechanism is
analogous in spirit to other zero-point-length/T-duality inspired regularization schemes
(e.g.\ Refs.~\cite{Padmanabhan:1996ap,Fontanini:2005ik,Nicolini:2022rlz} and the discussion around Refs.~\cite{Jusufi:2024dtr}).
 
 In the near future, we plan to further investigate possible signatures of the black hole solutions presented in this work. These include signatures of small primordial black holes and possible hollow structures \cite{Dai:2024guo}, as well as those of supermassive black holes, such as quasinormal modes (QNMs), shadow analysis, weak and strong lensing, thermodynamic properties, and other related phenomena. Another important extension of the present work would be the computation of grey-body factors associated with the modified black hole geometry. These factors encode deviations from the pure blackbody spectrum of Hawking radiation due to scattering in the effective potential surrounding the black hole. Studying grey-body factors in this framework could provide further insight into the evaporation process and the observational signatures of such regular black holes. We leave this analysis for future work.

\section*{Acknowledgment} K.J. and D.S. would like to thank Tian Zhou for insightful comments. D.S. acknowledges a CSU Fresno RSCA award and support from the Frank Sutton Research Fund.

\bibliography{ref}

\end{document}